\documentclass[twocolumn,trackchanges,twocolappendix]{aastex631}

\newcommand{\bootes}{Bo{\"o}tes}
\usepackage{CJKutf8}

\shorttitle{Specter}
\shortauthors{Chandra et al.}

\graphicspath{{./}{fig/}}

\begin{document}

\title{A Ghost in \bootes{}: The Least Luminous Disrupted Dwarf Galaxy}

\author[0000-0002-0572-8012]{Vedant~Chandra}
\affiliation{Center for Astrophysics $\mid$ Harvard \& Smithsonian, 60 Garden St, Cambridge, MA 02138, USA}

\author[0000-0002-1590-8551]{Charlie~Conroy}
\affiliation{Center for Astrophysics $\mid$ Harvard \& Smithsonian, 60 Garden St, Cambridge, MA 02138, USA}

\author[0000-0003-2352-3202]{Nelson~Caldwell}
\affiliation{Center for Astrophysics $\mid$ Harvard \& Smithsonian, 60 Garden St, Cambridge, MA 02138, USA}

\author[0000-0002-7846-9787]{Ana~Bonaca}
\affiliation{The Observatories of the Carnegie Institution for Science, 813 Santa Barbara Street, Pasadena, CA 91101, USA}

\author[0000-0003-3997-5705]{Rohan~P.~Naidu}
\affiliation{Center for Astrophysics $\mid$ Harvard \& Smithsonian, 60 Garden St, Cambridge, MA 02138, USA}

\author[0000-0002-5177-727X]{Dennis~Zaritsky}
\affiliation{Steward Observatory and Department of Astronomy, University of Arizona, Tucson, AZ 85721, USA}

\author[0000-0002-1617-8917]{Phillip~A.~Cargile}
\affiliation{Center for Astrophysics $\mid$ Harvard \& Smithsonian, 60 Garden St, Cambridge, MA 02138, USA}

\author[0000-0002-6800-5778]{Jiwon~Jesse~Han}
\affiliation{Center for Astrophysics $\mid$ Harvard \& Smithsonian, 60 Garden St, Cambridge, MA 02138, USA}

\author[0000-0002-9280-7594]{Benjamin~D.~Johnson}
\affiliation{Center for Astrophysics $\mid$ Harvard \& Smithsonian, 60 Garden St, Cambridge, MA 02138, USA}

\author[0000-0003-2573-9832]{Joshua S. Speagle (\begin{CJK*}{UTF8}{gbsn}沈佳士\ignorespacesafterend\end{CJK*})}
\affiliation{David A. Dunlap Department of Astronomy \& Astrophysics, University of Toronto, 50 St George Street, Toronto ON M5S 3H4, Canada}
\affiliation{Dunlap Institute for Astronomy \& Astrophysics, University of Toronto, 50 St George Street, Toronto, ON M5S 3H4, Canada}
\affiliation{Department of Statistical Sciences, University of Toronto, 100 St George St, Toronto, ON M5S 3G3, Canada}

\author[0000-0001-5082-9536]{Yuan-Sen~Ting \begin{CJK*}{UTF8}{gbsn}(丁源森)\end{CJK*}}
\affiliation{Research School of Astronomy \& Astrophysics, Australian National University, Cotter Road, Weston, ACT 2611, Australia}
\affiliation{School of Computing, Australian National University, Acton ACT 2601, Australia}

\author[0000-0002-0721-6715]{Turner~Woody}
\affiliation{Center for Astrophysics $\mid$ Harvard \& Smithsonian, 60 Garden St, Cambridge, MA 02138, USA}


\correspondingauthor{Vedant Chandra}
\email{vedant.chandra@cfa.harvard.edu}

\begin{abstract}

\noindent We report the discovery of Specter, a disrupted ultrafaint dwarf galaxy revealed by the H3 Spectroscopic Survey. We detected this structure via a pair of comoving metal-poor stars at a distance of 12.5~kpc, and further characterized it with \textit{Gaia} astrometry and follow-up spectroscopy. Specter is a $25^\circ \times 1^\circ$ stream of stars that is entirely invisible until strict kinematic cuts are applied to remove the Galactic foreground. The spectroscopic members suggest a stellar age $\tau \gtrsim 12$~Gyr and a mean metallicity $\langle\text{[Fe/H]}\rangle = -1.84_{-0.18}^{+0.16}$, with a significant intrinsic metallicity dispersion $\sigma_{ \text{[Fe/H]}} = 0.37_{-0.13}^{+0.21}$. We therefore argue that Specter is the disrupted remnant of an ancient dwarf galaxy. With an integrated luminosity $M_{\text{V}} \approx -2.6$, Specter is by far the least-luminous dwarf galaxy stream known. We estimate that dozens of similar streams are lurking below the detection threshold of current search techniques, and conclude that spectroscopic surveys offer a novel means to identify extremely low surface brightness structures.

\end{abstract}


\section{Introduction} \label{sec:intro}

In the modern cosmological paradigm, galaxies form `bottom-up', with smaller galaxies coalescing and merging over cosmic time to assemble larger galaxies \citep[e.g.,][]{Press1974,White1978,Bullock2005}. 
Our own Milky Way continues growing to this day, as evidenced by its rich system of dwarf galaxy satellites \citep[e.g.,][]{Mateo1998,Simon2019}, and recent discoveries of phase-mixed accreted debris throughout the Galaxy \citep[e.g.,][]{Helmi2018,Belokurov2018a,Naidu2020,Malhan2022a}.
Over the past several decades, deep photometric surveys have revealed a plethora of ultrafaint dwarf galaxies ($M_{\text{V}} \gtrsim -7.7$, hereafter `dwarfs') surrounding the Milky Way \citep[e.g.,][]{Willman2005a,Willman2005b,Belokurov2007,Koposov2015a,Simon2019,Drlica-Wagner2020}. 
These surveys have likewise unearthed a patchwork of ex-situ stellar streams encircling the Galaxy, tidally stretched remnants of past satellites and their globular clusters \citep[e.g.,][]{Majewski2003,Belokurov2007a,Grillmair2009,Shipp2018,Bonaca2020b}. 
Although these emissaries are fascinating in their own right due to their extragalactic formation, they also teach us about the distribution of our own Galaxy's stellar and dark matter \citep[e.g.,][]{Murali1999,Eyre2011,Bonaca2014,Bovy2016b,Bonaca2018,Malhan2021a,Nibauer2022}. 


The key limitation of photometrically detecting nearby dwarfs and streams is the projected stellar surface density, since this sets the contrast between a given structure and the Milky Way foreground.
While these systems are all resolved into individual stars, this is usually expressed in terms of the effective surface brightness (SB) of a given density of stars. 
There exists a well-known SB frontier of $\mu \sim 31$~mag\,arcsec$^{-2}$ for intact dwarfs found with deep surveys like the Sloan Digital Sky Survey and Dark Energy Survey \citep{Koposov2008,Walsh2009,Drlica-Wagner2020}. 
It remains an open question whether the current SB frontier is purely an observational artefact, and if there exists a hidden population of `stealth' galaxies invisible to photometric star-counting searches \citep{Tollerud2008,Bullock2010,Munoz2018b}. These systems probe the lowest mass scales of galaxy formation, and are vital to our understanding of dark matter subhalos, structure formation, and even reionization \citep[e.g.,][]{Klypin1999,Moore2006,Jethwa2018,Hayashi2022}. 

Recent works, armed with proper motions from \textit{Gaia} \citep{GaiaCollaboration2021,Lindegren2021}, have pushed the SB frontier by employing kinematic filtering to peer through the Galactic foreground and detect more diffuse intact dwarfs \citep{Torrealba2019}.
Incorporating kinematic information has likewise pushed the frontier to detect fainter and more diffuse stellar streams \citep[e.g.,][]{Malhan2018,Ibata2019,Yuan2020,Tenachi2022,Oria2022}. 
The best-characterized streams with dwarf progenitors have typical luminosities $M_{\text{V}} \sim -5.5$, whereas those with globular cluster (GC) progenitors tend to be narrower and have $M_{\text{V}} \sim -3.5$ \citep{Shipp2018,Ji2020}. 
Although width can roughly indicate whether a stream had a dwarf or GC progenitor, a more conclusive test is the detection of an intrinsic metallicity dispersion, indicative of multiple generations of star formation \citep{Gilmore1991,Willman2012}. 

Here we present the discovery of Specter, a diffuse stellar stream revealed by the H3 Survey \citep{Conroy2019b}. 
We measure Specter's structural and chemical properties, and argue that it is the faintest disrupted dwarf galaxy known to date. 
We suggest that Specter is representative of a large population of diffuse dwarf streams that are invisible to traditional search techniques. 
We describe our search for structures in the H3 Survey in $\S$\ref{sec:data}, and characterize the morphology and stellar population of Specter in $\S$\ref{sec:specter}. 
Finally, we discuss the implications of our findings in $\S$\ref{sec:discuss}. 

\section{Data and Discovery}\label{sec:data}

\subsection{Search for Structures in H3}

The H3 Survey \citep{Conroy2019b} has been collecting high-resolution $R \approx 32\,000$ spectra for parallax-selected halo stars since 2017, using the Hectochelle instrument on the MMT \citep{Szentgyorgyi2011}. 
To-date, H3 has observed $ 240\,000$ stars with $\pi \lesssim 0.4$~mas and $15 \lesssim r \lesssim 18.5$, employing a sparse tiling strategy across the entire $\mid b \mid > 30^\circ$ (off-plane) and $\delta > -20^\circ$ (visible from the MMT) sky. 
H3 spectra cover the Mg\,{\it b} triplet from $5150-5300$\,\AA, and are analyzed with the full-spectrum \texttt{MINESweeper} pipeline to deliver radial velocities, spectrophotometric distances, metallicities, and $\alpha$-abundances \citep{Cargile2020}. 

For the present study, our initial goal was to detect low-luminosity intact dwarfs, whose masses might be so low that only a few stars would be bright enough to appear in the combined \textit{Gaia} and H3 catalogs.  
Such sparse populations would not normally pass a photometric detection threshold, but we expect the addition of chemistry and 3D kinematics to significantly enhance the likelihood of physical association.
We emphasize the importance of H3's $\approx 1$~km\,s$^{-1}$ radial velocity precision, which helps us isolate truly cold structures without being dominated by measurement noise. 
We thus performed a search for cospatial and comoving pairs of stars in the H3 data collected as of April~2022, leveraging the power of \textit{Gaia} DR3 astrometry \citep{Lindegren2021} and H3 radial velocities and stellar parameters to find comoving stars.
Specifically, we started with the \texttt{rcat\_V4.0.5.d20220422\_MSG} H3 catalog. 
To reduce contamination from nearby disk stars, we first selected stars with distance $d > 10$~kpc and [Fe/H]~$< -1.5$, with spectral signal-to-noise greater than 2. 
We also removed stars belonging to known cold structures like globular clusters, and removed plausible members of the Sagittarius stream using the angular momentum cuts described in \cite{Johnson2020a}. 
For each pairwise combination of stars in this sample, we retained dyads that satisfied the following criteria: 
\texttt{
    ($\sqrt{\Delta \alpha^2 + \Delta \delta^2}$ < 3$^\circ$) AND ($\sqrt{\Delta \mu_{\alpha}^2 / \sigma_{\mu_\alpha}^2 + \Delta \mu_{\delta}^2 / \sigma_{\mu_\delta}^2}$ < 2) AND ($\Delta v_{r,h}$ < 10 km\,s$^{-1}$)
    }. 

This search yielded 6 groups of comoving stars containing $2-3$~members each. 
We manually vetted the groups and compared them to known dwarf galaxies, globular clusters, and stellar streams \citep{McConnachie2012,Baumgardt2020,Mateu2022}. 
Our prior expectation was that the majority of pairs would reside in known streams, since stellar streams span tens of degrees on the sky and have a significantly enhanced probability of hosting comoving stars within a few degrees of each other. 
Groups~\#1 and \#2 could securely be associated with the Cetus stream based on their sky position and kinematics \citep{Newberg2009,Yuan2022}. 
Group~\#3 was likewise strongly associated with the Orphan stream \citep{Belokurov2007a,Koposov2019}. 
Groups \#4 and \#5 could not be immediately associated with known structures, but did not stand out as significant new structures either. Specifically, there was no strong spatial overdensity around these structures after kinematic and color-magnitude filtering was applied.
Groups \#4 and \#5 may be associated with the Sagittarius stream since they are only $\approx 10$~degrees off the Sagittarius orbital plane, and were perhaps missed by our angular momentum cuts to mask out Sagittarius. 
We regard their status as uncertain. Group~\#6 could not be associated with any known structures, and is the subject of this paper. 

\subsection{A Ghost in \bootes}\label{sec:data:ghost}

\begin{figure}
    \centering
    \includegraphics[width=\columnwidth]{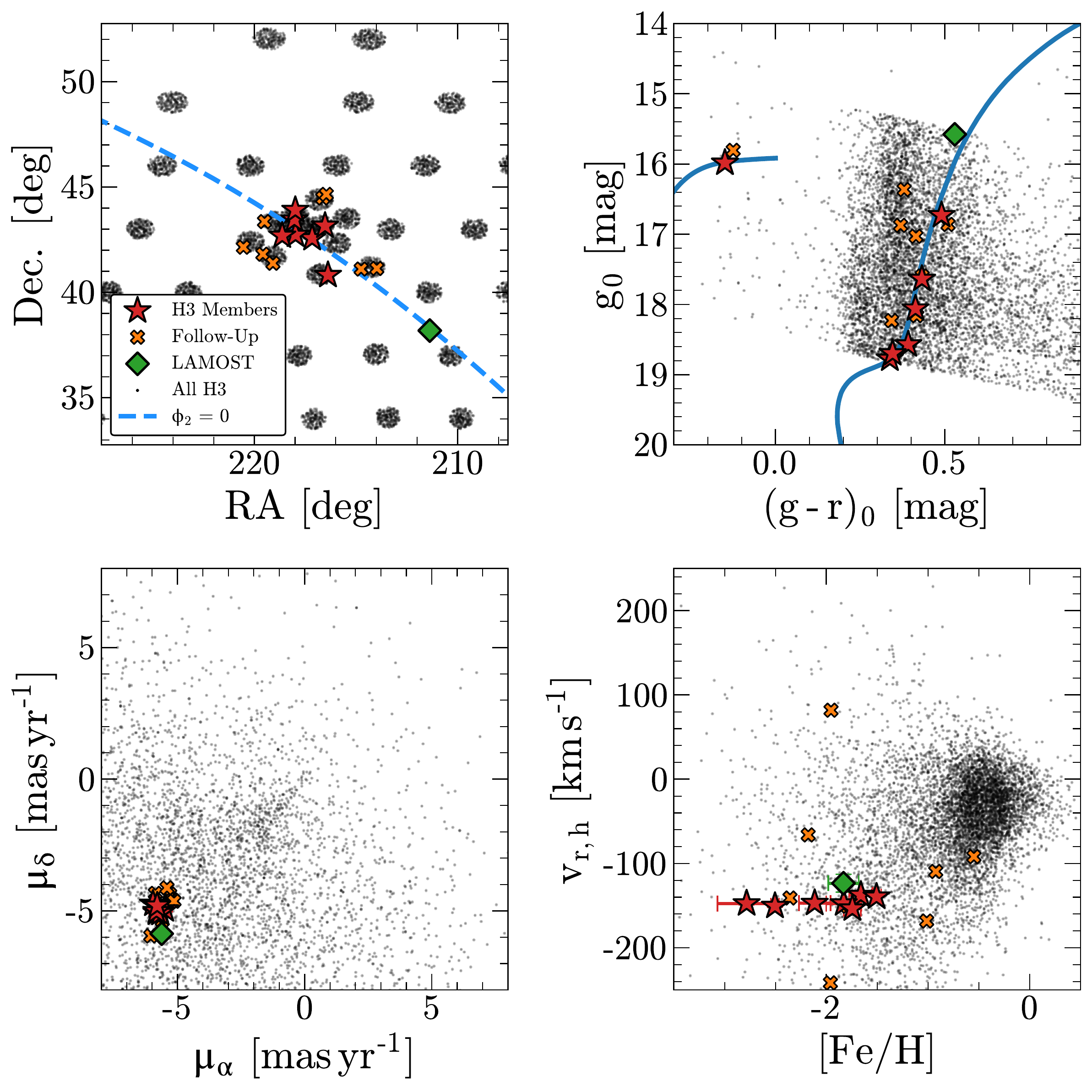}
    \caption{Identification of spectroscopic members from H3 Survey data. The black points denote all H3 objects in the $10^\circ \times 10^\circ$ field of view shown in the top left panel, orange crosses denote PM-selected stars that were specifically followed up for this work, and the red points are our final Specter members. We show the $\phi_2=0$ line that defines our great circle coordinate frame. In the top right CMD, we overlay a [Fe/H]$ = -1.8$, 13~Gyr MIST isochrone, as well as the BHB ridgeline from \cite{Deason2011}. Our members lie on the same isochrone, have near-identical proper motions and radial velocities, and are systematically metal-poor with spectroscopic [Fe/H]$ < -1.5$. We also show the tentative LAMOST member described in $\S$~\ref{sec:data:seglam}.}
    \label{fig:h3_summary}
\end{figure}

Our search identified a pair of low-metallicity RGB stars (Group $\#6$ above) at $d \approx 12.5$~kpc that did not appear to be associated with any known structures, with near-identical proper motions and radial velocities. 
Motivated by this, we observed an additional 11 H3 fields around this pair, specifically targeting stars with similar proper motions that plausibly lay on the same isochrone. 
This yielded seven `member' stars within two degrees of each other that lie on the same isochrone and have exceptionally similar proper motions and radial velocities (Figure~\ref{fig:h3_summary}, Table~\ref{tab:spec_mem}). 
As Figure~\ref{fig:h3_summary} illustrates, several of the follow-up members would not have fallen in the standard H3 selection because they are outside the sparse on-sky footprint of the regular survey.
We elaborate on the detection significance of this structure in Appendix~\ref{sec:detsig}, and show the H3 spectra in Appendix~\ref{sec:h3spec}.

We use the member blue horizontal branch (BHB) star and the empirical ridgeline from \cite{Deason2011} to derive a distance $d = 12.5 \pm 0.5$~kpc, adopting a 0.1~mag uncertainty in the BHB ridgeline. 
The other member stars have spectrophotometric distances from \texttt{MINESweeper} that are consistent with this value at the $1\sigma$ level, with an error-weighted mean distance $\approx 13.3 \pm 0.5$~kpc, albeit with a larger systematic uncertainty. The parallax distribution of these stars is likewise consistent with this distance, with an error-weighted mean parallax corresponding to a distance $\approx 15_{-3}^{+6}$~kpc after applying the zeropoint correction \citep{Lindegren2018}. 

To investigate if these stars are part of a larger structure like a dwarf galaxy or stellar stream, we queried \textit{Gaia} DR3 for all stars in this region of the sky. 
We removed foreground contaminants with a stringent parallax selection of $\pi < 0.2$~mas and cross-matched our sample with photometry from Pan-STARRS \citep{Chambers2016}. 
We use Pan-STARRS optical photometry because it is deeper than \textit{Gaia}, and \textit{Gaia} $\text{G}_{\text{BP}}$ colors are known to suffer from systematic issues for faint red sources \citep{Arenou2018}.
Our results are qualitatively unchanged if we instead use the \textit{Gaia} $\text{G}$, $\text{G}_{\text{BP}}$, $\text{G}_{\text{RP}}$ photometric system, albeit the CMD colors are noisier at the faint end.
We de-reddened the Pan-STARRS photometry using dust maps from \cite{Schlegel1998}, re-normalized by \cite{Schlafly2011}. 
Since we only desire stars with reliable proper motion measurements, we adopt a limiting magnitude of $G = 20.5$.

\begin{deluxetable*}{cccccccccc}
\label{tab:spec_mem}
\tablewidth{1\textwidth} 
\tablecaption{Specter Members with Spectroscopy.}
\tablehead{\colhead{\textit{Gaia} Source ID} & \colhead{RA} & \colhead{Dec.} & \colhead{PS $g$} & \colhead{SNR} & \colhead{$v_{r,h}$} & \colhead{$T_{\text{eff}}$} & \colhead{$\log{g}$} & \colhead{[Fe/H]} & \colhead{[$\alpha$/Fe]}\\ \colhead{DR3} & \colhead{deg} & \colhead{deg} & \colhead{mag} & \colhead{px$^{-1}$} & \colhead{$\text{km\,s}^{-1}$} & \colhead{K} & \colhead{$\log{\left[\text{cm}\,\text{s}^{-2}\right]}$} & \colhead{dex} & \colhead{dex}}
\startdata
1492503132521811712 & 216.527 & 43.155 & 16.0 & $14.6$ & $-150.5 \pm 1.1$ & $8270 \pm 70$ & $\cdots$\tablenotemark{a} & $\cdots$ & $\cdots$ \\
1491676299777912064 & 217.162 & 42.568 & 16.8 & $13.3$ & $-139.4 \pm 0.2$ & $5150 \pm 20$ & $2.6 \pm 0.1$ & $-1.51 \pm 0.05$ & $0.0 \pm 0.1$ \\
1494105876877711104 & 217.989 & 43.907 & 17.7 & $6.8$ & $-153.6 \pm 0.3$ & $5200 \pm 30$ & $2.8 \pm 0.1$ & $-1.74 \pm 0.08$ & $0.3 \pm 0.1$ \\
1491186982743561088 & 216.383 & 40.817 & 18.1 & $4.4$ & $-137.4 \pm 0.5$ & $5350 \pm 30$ & $3.2 \pm 0.1$ & $-1.66 \pm 0.12$ & $0.3 \pm 0.1$ \\
1491737528831174912 & 217.978 & 42.729 & 18.6 & $3.7$ & $-147.0 \pm 0.8$ & $5410 \pm 30$ & $3.2 \pm 0.1$ & $-2.11 \pm 0.16$ & $0.3 \pm 0.2$ \\
1493316732469979136 & 218.035 & 43.452 & 18.7 & $2.4$ & $-147.5 \pm 1.8$ & $5520 \pm 40$ & $3.2 \pm 0.1$ & $-2.78 \pm 0.29$ & $0.2 \pm 0.2$ \\
1493037765754139136 & 218.634 & 42.686 & 18.8 & $2.8$ & $-147.2 \pm 1.1$ & $5670 \pm 40$ & $3.6 \pm 0.1$ & $-1.72 \pm 0.18$ & $0.3 \pm 0.2$ \\
\tableline
1483861280364752512 & 211.377 & 38.185 & 15.6 & $\cdots$ & $-123 \pm 10$ & $5100 \pm 100$ & $2.3 \pm 0.3$ & $-1.8 \pm 0.2$ & $0.2 \pm 0.1$
\enddata
\tablecomments{The first seven stars are from the H3 Survey, with stellar parameters estimated by \texttt{MINESweeper}. The last star is the tentative LAMOST member, with stellar parameters from \cite{Xiang2019}. \textbf{Listed uncertainties are statistical only.} Since there are $\approx 2$ pixels per resolution element, the SNR per resolution element is $\approx \sqrt{2}$ times larger than the values listed here.}
\tablenotetext{a}{The BHB is too hot to derive reliable parameters beyond temperature and radial velocity with the H3 spectrum, so we omit listing those parameters here.}
\end{deluxetable*}

We compute the mean orbit of our H3 members with \texttt{gala} \citep{gala,adrian_price_whelan_2020_4159870}, assuming the default \texttt{MilkyWayPotential} \citep{Bovy2015}. 
We use this orbit to define a great circle coordinate system $(\phi_1, \phi_2)$ with a pole at $(\alpha,\delta) = (269.76^\circ, -33.55^\circ)$, and set the zeropoint $(\phi_1, \phi_2) = (0,0)$ at $(\alpha,\delta) = (217.53^\circ, -42.76^\circ)$. 
For each star in our \textit{Gaia} sample, we assign a distance based on the nearest on-sky point on the orbit \citep[e.g.,][]{Price-Whelan2018b,Bonaca2019a}. 
We transform the DR3 proper motions to the great circle frame and correct them for the solar reflex motion \citep{Schonrich2010}. 

\begin{figure}
    \centering
    \includegraphics[width=\columnwidth]{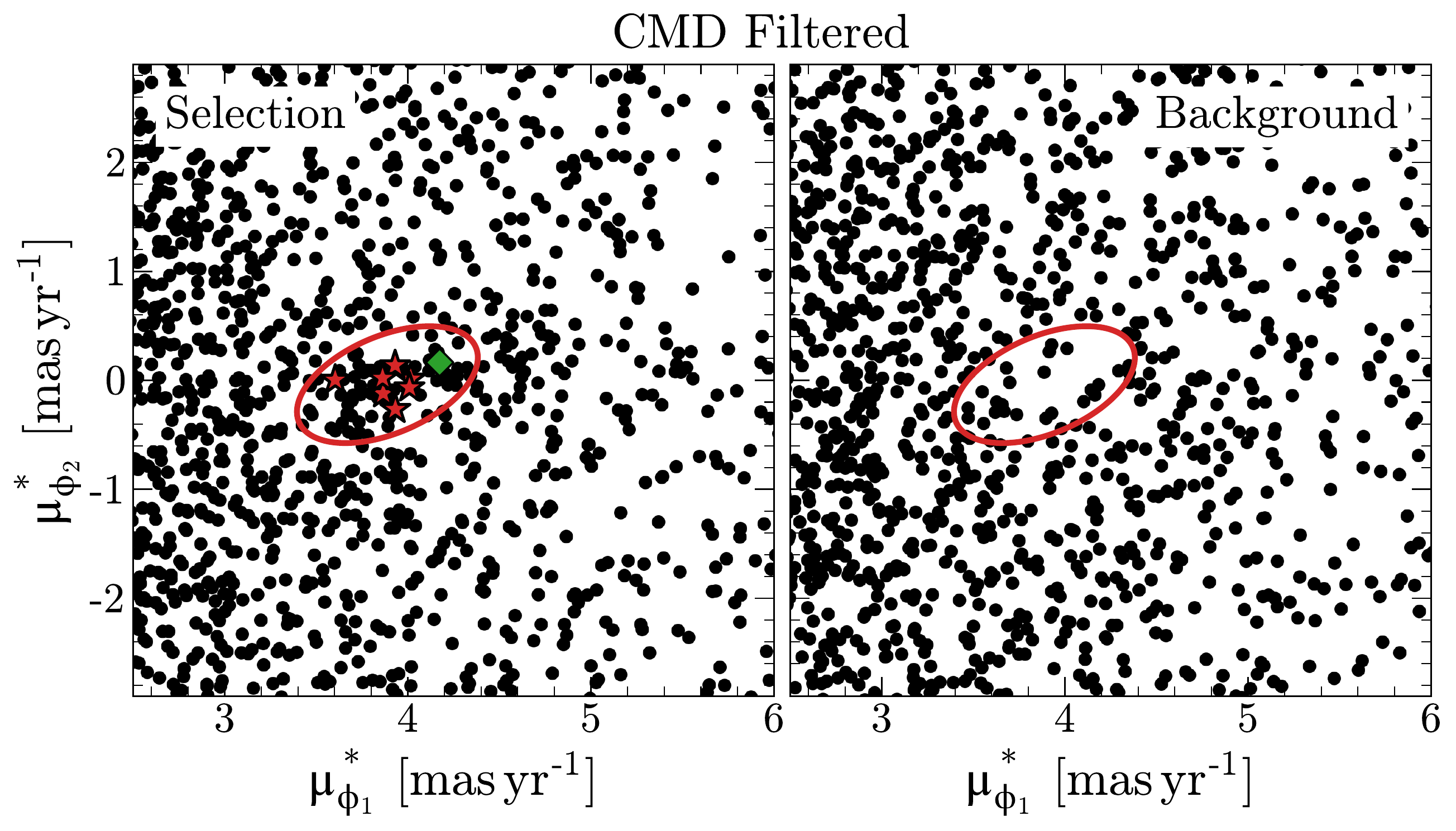}
    \includegraphics[width=\columnwidth]{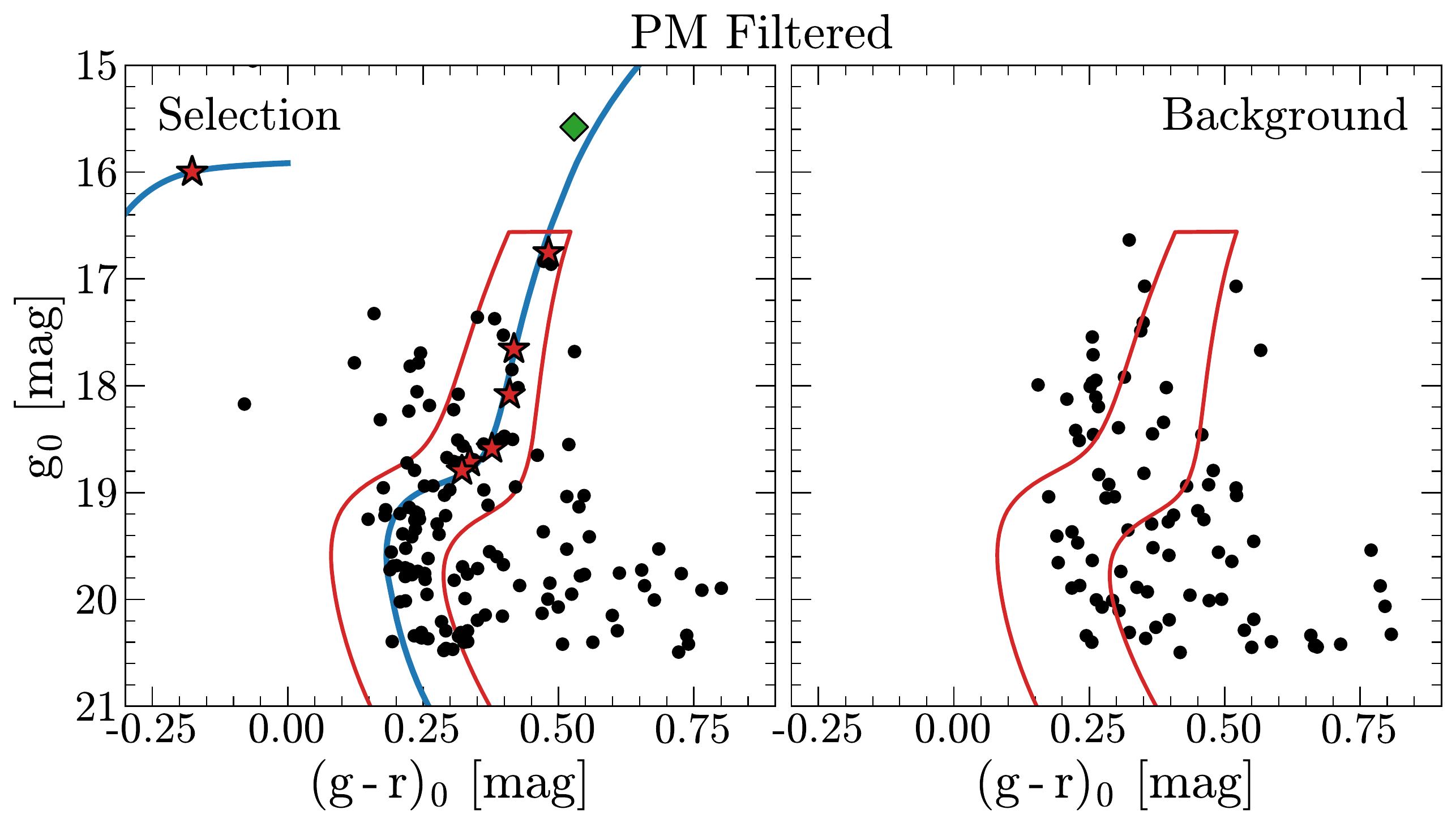}
    \includegraphics[width=\columnwidth]{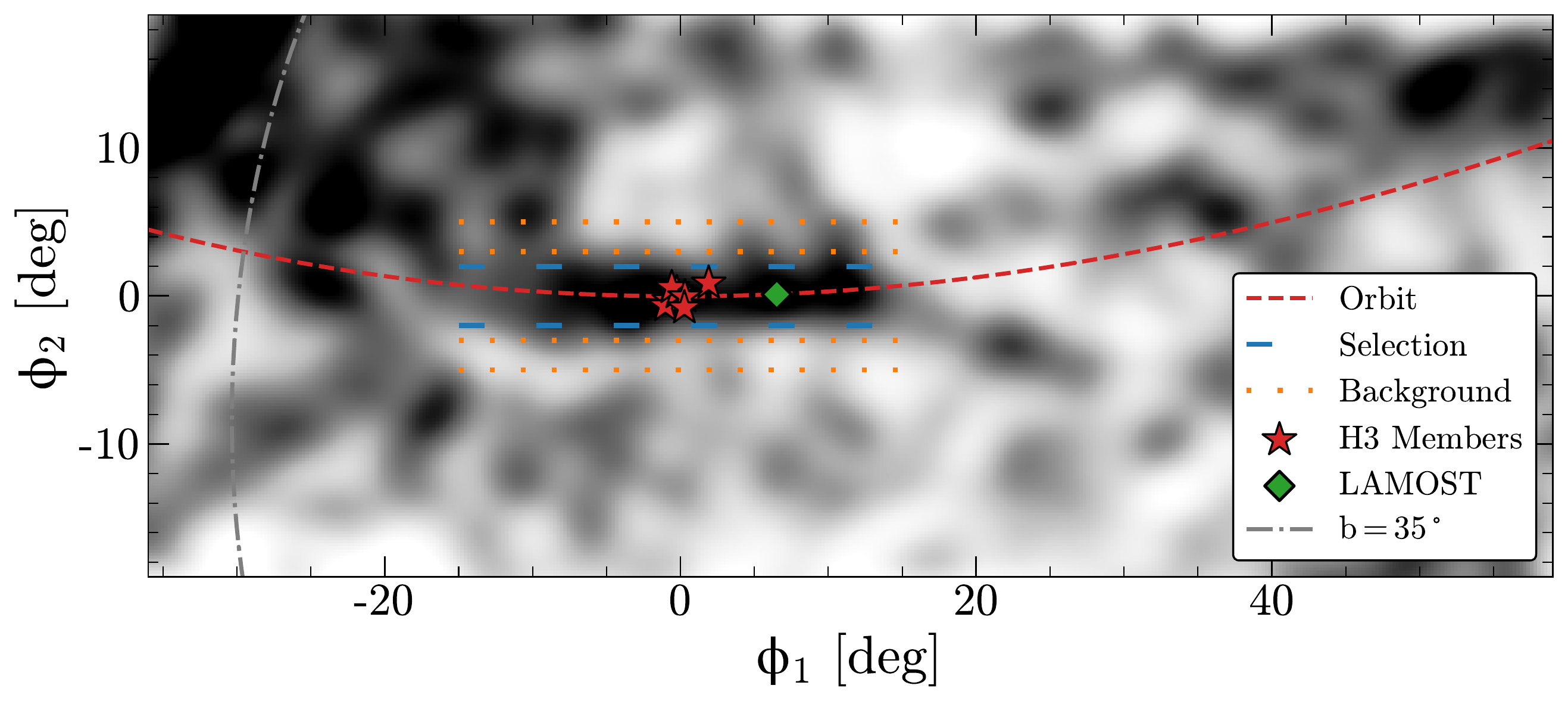}
    \caption{Summary of our selection procedure to reveal Specter in \textit{Gaia} DR3 data. Top: Reflex-corrected proper motions in our transformed great circle coordinates, after applying an isochrone filter to select stars at $\approx 12.5$~kpc. Our proper motion selection is shown with a red ellipse, and is centred on the mean proper motion of the H3 members. Middle: CMD of stars that lie in the proper motion ellipse shown in the top panel, without any CMD filtering applied. We overlay a fiducial 13~Gyr MIST isochrone with [Fe/H]$=-1.8$ and [$\alpha$/Fe]$=+0.3$. Bottom: Spatial density of stars that pass both the PM and CMD filters, smoothed with a $1.5^\circ$ Gaussian kernel. We overlay the spectroscopic members and their predicted orbit.}
    \label{fig:selection}
\end{figure}

Figure~\ref{fig:selection} illustrates the transformed proper motion and color-magnitude spaces for a region close to our H3 members, and an equal-area background region. 
There is a clear overdensity of stars with proper motions similar to our spectroscopic members (top panel). 
Upon filtering for stars with these proper motions --- shown by the $\approx 1$~mas\,yr$^{-1}$ red ellipse centered at $(\mu_{\phi_1}^\ast, \mu_{\phi_2}^\ast) \approx (3.8, 0)$~mas\,yr$^{-1}$ --- a coherent stellar population reveals itself on the color-magnitude diagram (middle panel). 
Although the number of stars is small, the CMD overdensity is significant at the $\approx 7\sigma$ level compared to the equal-area background region based on Poisson statistics.
After further selecting stars that lie within 0.1~mag (in color) of a fiducial [Fe/H]$ = -1.8$, [$\alpha$/Fe]$ = +0.3$, 13~Gyr MIST isochrone \citep{Dotter2016,Choi2016}, we illustrate their spatial density in the bottom panel of Figure~\ref{fig:selection}. 
An elongated overdensity spanning $\approx 25^\circ$ appears into view, surrounding our spectroscopic members and tracing their predicted orbit. 
This extended stellar system is clearly a disrupted stream, not a gravitationally bound dwarf galaxy. 
We name this structure \textit{Specter} to reflect its ephemeral nature, and to highlight the role of spectroscopy in its discovery\footnote{We follow the tradition of naming stellar streams after bodies of water, namely the Specter Rapids in Arizona, the state where the MMT is located.}. 

The central region of Specter was coincidentally observed as a part of the Hyper Suprime-Cam Subaru Strategic Program in the WIDE field \citep{Aihara2018,Aihara2019,Aihara2022}.
This field covers only a roughly $2^\circ \times 2^\circ$ portion of the stream. 
We searched this dataset for evidence of main sequence Specter members between $19 \lesssim r \lesssim 23$, after performing a standard star-galaxy separation with $griz$ colors. 
However, we found the background-subtracted color-magnitude diagrams to be quite noisy and lacking clear evidence of a main sequence. 
While this problem is exacerbated by the restricted region covered by the HSC photometry, we mainly attribute it to how elusive Specter is without kinematic filtering. 
In the absence of proper motion information, the stellar population of Specter is drowned out by the Galactic foreground.

\subsection{Search for Members in SEGUE and LAMOST}\label{sec:data:seglam}

To investigate whether Specter has additional plausible members in existing spectroscopic datasets, we performed a search in the Sloan Extension for Galactic Understanding and Exploration (SEGUE;  \citealt{Yanny2009,Eisenstein2011,Alam2015}) and Large Sky Area Multi-Object Fibre Spectroscopic Telescope (LAMOST; \citealt{Cui2012,Zhao2012}) catalogs. 
After performing basic cleanliness cuts to remove bad fits and stars with [Fe/H]$> -1$, we cross-matched these catalogs to \textit{Gaia} DR3 astrometry around Specter.
We performed a coordinate transformation and reflex-correction as described in $\S$\ref{sec:data:ghost}, and applied the proper motion selection shown in Figure~\ref{fig:selection}. 
As a function of the $\phi_1$ coordinate, we then selected stars that lie within $\phi_2 \pm 3^\circ$ and $v_{r,h} \pm 15$~km\,s$^{-1}$ of our predicted orbit for Specter. 
This search yielded zero stars in SEGUE and one star in LAMOST, which we display in Figure~\ref{fig:h3_summary}. 
Relaxing the kinematic selections by a factor of two does not produce more members that plausibly lie on the same isochrone as our H3 members. 

\textit{Gaia}\,DR3\,1483861280364752512 lies at $(\phi_1, \phi_2) \approx (6.5, 0.1)$, and has a reported metallicity [Fe/H]\,$=-1.83 \pm 0.15$ in the LAMOST catalog of \cite{Xiang2019}. 
This star matches our mean metallicity for Specter and lies within the $\phi_1 = \pm 10^\circ$ region in which we detect a significant overdensity of Specter stars in \textit{Gaia}.
We therefore tentatively assign it as a spectroscopic member of the stream and display it in Figure~\ref{fig:h3_summary}.
We conservatively do not include this star in our detailed analysis of the stream (e.g., to derive the velocity and metallicity dispersion) and instead use the homogeneous set of seven H3 members. 

Our search did not reveal more promising spectroscopic members out to $\phi_1 \pm 50^\circ$ around Specter in either SEGUE or LAMOST. 
Beyond this, uncertainties in the orbit (due to measurement errors and systematic uncertainties in the Milky Way potential) likely make it challenging to confidently associate stars with Specter.

\section{Specter}\label{sec:specter}

\subsection{Structural Parameters}

We estimate the spatial structure of Specter by fitting the $\phi_2$ distribution of stars in bins of $\phi_1$. 
We use the sample of stars shown in Figure~\ref{fig:selection} that pass our proper motion and isochrone selection. 
In each of 15 overlapping $\phi_1$ bins ($8^\circ$ wide and spaced $4^\circ$ apart), we model the $\phi_2$ distribution as a mixture of a linear background and a Gaussian component for the stream \citep[e.g.,][]{Bonaca2019a}. 
The free parameters are the mean and (log) width of the Gaussian component, the slope of the linear background, and the fraction of stars in the Gaussian component.
There are 98--186 stars in each bin.
We construct a likelihood and sample the parameter posterior distributions with \texttt{emcee} \citep{Foreman-Mackey2013,Foreman-Mackey2019}, assuming uniform priors on all parameters.
32 walkers are used to sample each bin's posterior for 1024 steps, discarding the first half as burn-in and taking the median estimate for each parameter. 

\begin{figure}
    \centering
    \includegraphics[width=\columnwidth]{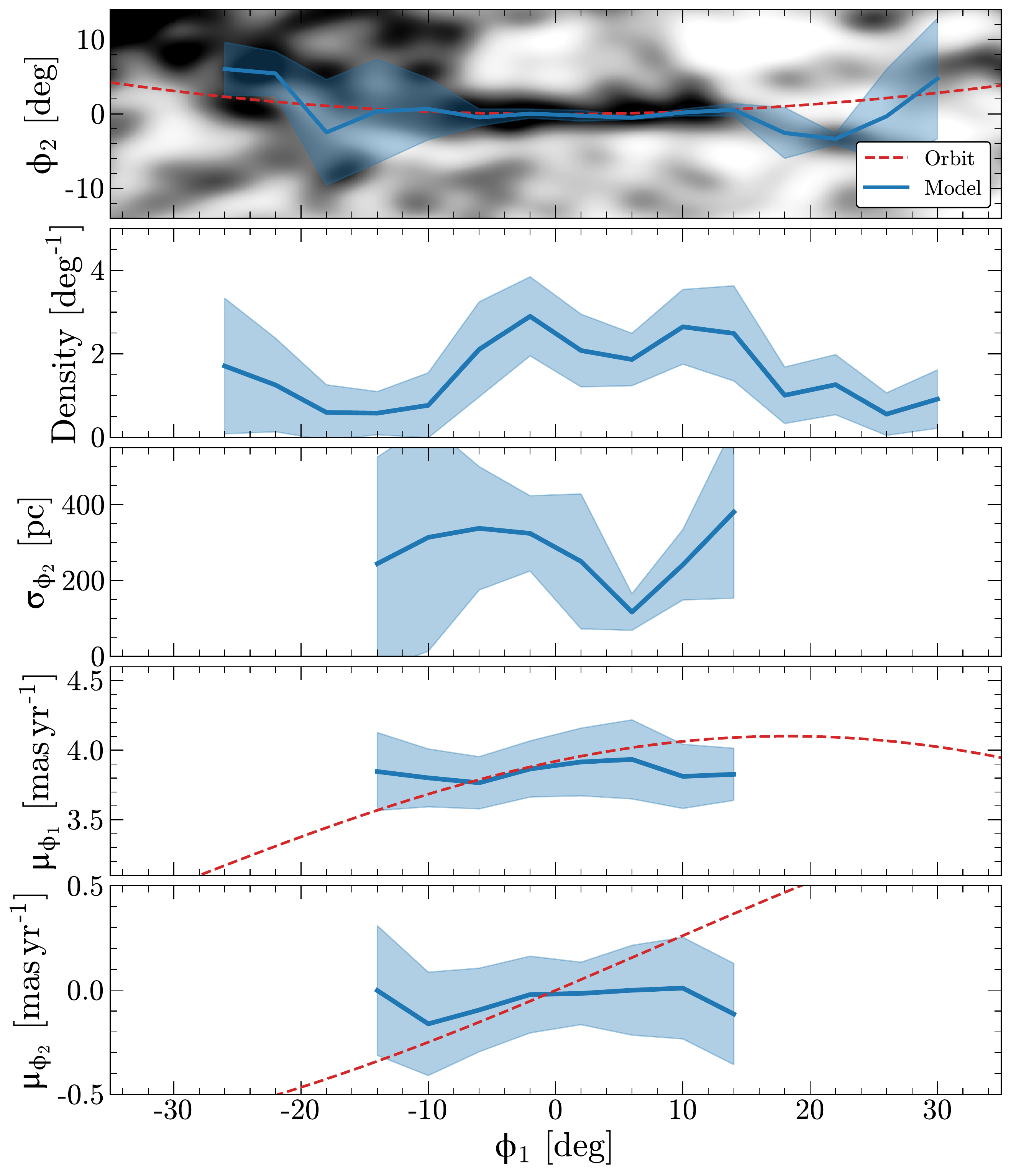}
    \caption{Mixture modeling the spatial structure of Specter. Panel 1 from the top: Observed spatial density of stars that pass our proper motion and isochrone selection. We overlay the fitted mean $\phi_2$ of the Gaussian component in each $\phi_1$ bin, along with the predicted orbit of the H3 member stars. Panel 2: The density of stars in the Gaussian stream component of each bin. Panel 3: The standard deviation of the Gaussian component, transformed to physical units using the distance predicted by the orbit model. Panel $4-5$: median and standard deviation of the proper motions of stars in the Gaussian component.}
    \label{fig:density_fit}
\end{figure}

Figure~\ref{fig:density_fit} summarizes our model fit across the stream.
Specter is well-constrained between $-10^\circ \lesssim \phi_1 \lesssim 15^\circ$, beyond which the stellar density rapidly falls below our detection limits.
The width and median proper motion measurements are shown in the bottom panels of Figure~\ref{fig:density_fit} for the well-constrained stream region. 
In this range, the width (Gaussian $\sigma$) is $\approx 1^\circ$, or $\approx 200$~pc using the adopted distance. 
Both the predicted and observed proper motion gradients are quite shallow in this region of the stream. 

\subsection{Stellar Population}\label{sec:stellpop}

Our spatial model produces membership probabilities  $P_{\text{mem}} = P_{\text{G}} / (P_{\text{G}} + P_{\text{bg}})$ where $P_{\text{G}}$ and $P_{\text{bg}}$ denote a star's likelihood of being in the Gaussian or background components respectively. 
Again, we only utilize stars that pass the CMD and PM cuts shown in Figure~\ref{fig:selection}. 
The total number of stars above our limiting magnitude of $G = 20.5$ is approximately $\sum P_{\text{mem}} \approx 100$ in the inner $-10^\circ \lesssim \phi_1 \lesssim 15^\circ$ region.
Using the formalism of \cite{Martin2008}, adopting a [Fe/H]$ = -1.8$, 13~Gyr MIST isochrone and a \cite{Kroupa2001} IMF, this implies an integrated luminosity $M_{\text{V}} \approx -2.6 \pm 0.5$. 
If we simply add the fluxes of stars with $P_{\text{mem}} > 0.8$ and only use the \cite{Martin2008} formalism to integrate the unseen stellar flux, we derive a similar value $M_{\text{V}} \approx -2.6$.
Summing the IMF-weighted MIST isochrone, the total stellar mass is consequently $M_\ast \approx 2000 \pm 500\,M_\odot$  ($\log{M_\ast / M_\odot} \sim 3.3$). 
Assuming that 68\% of the luminosity is contained within the derived Gaussian width of the stream, the implied surface brightness of Specter is $\mu \sim 34$~mag\,arcsec$^{-2}$. 

\begin{figure}
    \centering
    \includegraphics[width=\columnwidth]{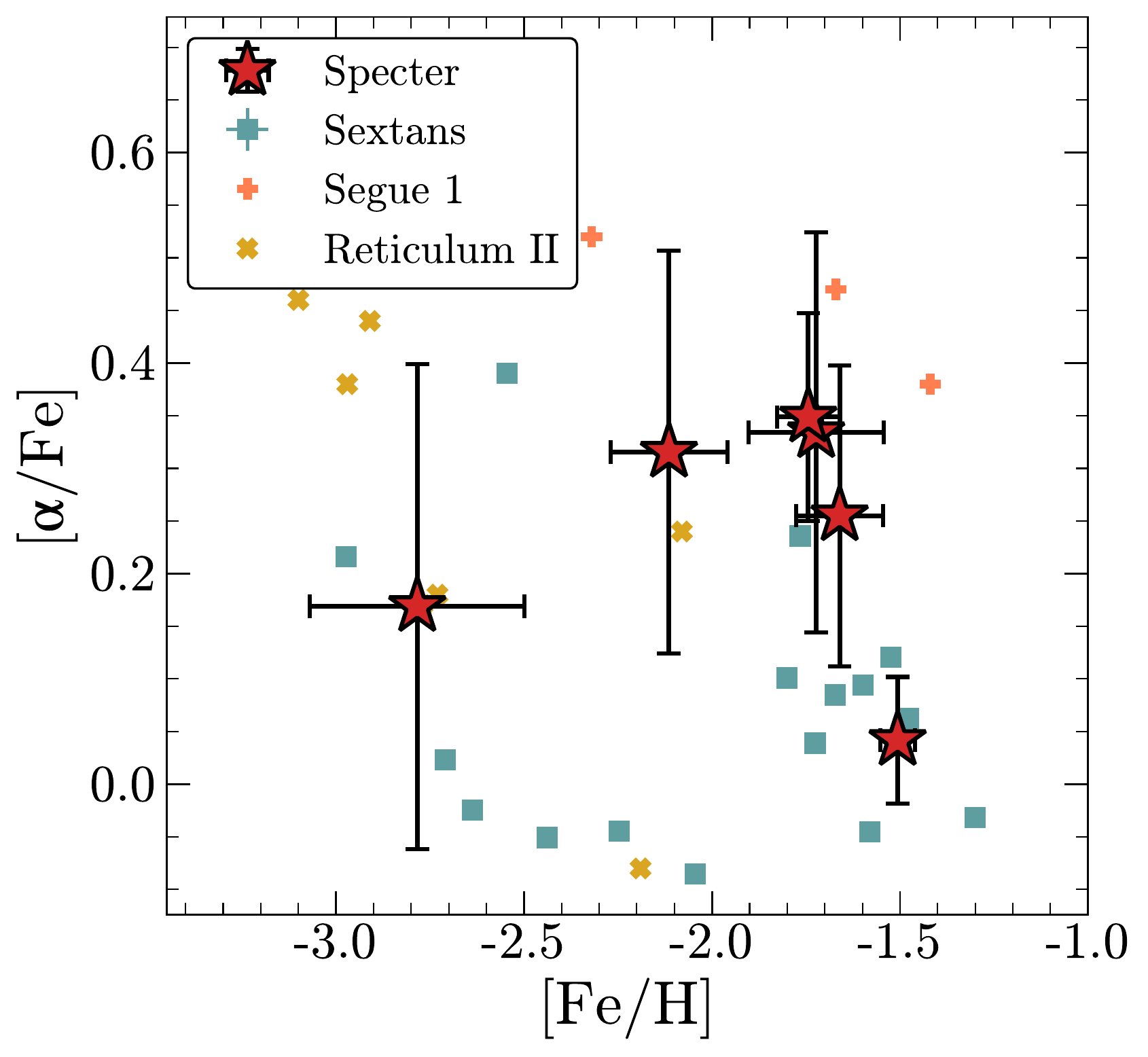}
    \caption{Tinsley-Wallerstein diagram of Specter members using measurements from H3, excluding the BHB star. We overlay abundance measurements for ultra-faint dwarf galaxies with similar luminosities \citep{Frebel2014,Ji2016}. We also show stars in the Sextans dwarf galaxy observed by H3, which should be on an identical metallicity scale to the Specter stars.}
    \label{fig:feh_afe}
\end{figure}

Figure~\ref{fig:feh_afe} illustrates our H3 spectroscopic members in chemical space. 
Although the sample is small, metal-poor members are systematically $\alpha$-enhanced with an anti-correlation between [$\alpha$/Fe] and [Fe/H], suggesting an extended star formation history in Specter's progenitor wherein the gas was gradually enriched by Type\,Ia supernovae \citep{Gilmore1989,Kirby2011}. 
To further investigate the intrinsic metallicity distribution of Specter, we fit a Gaussian distribution to the metallicities using a maximum-likelihood approach \citep[e.g.,][]{Walker2006}. 
We sample the posterior distribution of the Gaussian parameters with \texttt{emcee} and assume a uniform prior for the log-dispersion. 
Specter has a mean metallicity $\langle [\text{Fe/H}] \rangle = -1.84_{-0.18}^{+0.16}$, and an intrinsic metallicity dispersion $\sigma_{[\text{Fe/H}]} = 0.37_{-0.13}^{+0.21}$ dex. 
By performing a similar fit to the [$\alpha$/Fe] measurements, we derive $\langle [\text{$\alpha$/Fe}] \rangle =+0.22_{-0.07}^{+0.07}$, with an unresolved intrinsic dispersion. 
The metallicity dispersion becomes unresolved if we exclude the most metal-poor star, but we have no reason to doubt its membership or metallicity. 
This star's spectrum is discussed in more detail in Appendix~\ref{sec:h3spec}, and detailed validation of H3's metallicity scale is presented in \cite{Cargile2020}.

\begin{deluxetable}{lcr}\label{tab:params}
\tablewidth{\columnwidth}
\tablecaption{Measured Properties of Specter.}
\tablehead{
\colhead{Parameter} & \colhead{Value} & \colhead{Unit}}
\startdata
\sidehead{Overview}
$d_\odot$ & $12.5 \pm 0.5$ & kpc \\
$M_V$ & $-2.6 \pm 0.5$ & mag \\
$M_\ast$ & $\approx 2000 \pm 500$ & $M_\odot$ \\ 
Length & $\approx 5.5$ & kpc \\
Width & $\approx 200$ & pc \\
\tableline
\sidehead{H3 Members}
$\langle \mu_{\text{RA}} \rangle$ & $-5.75 \pm 0.02$  & mas\,yr$^{-1}$ \\
$\langle \mu_{\text{DEC}} \rangle$ & $-5.00 \pm 0.02$  & mas\,yr$^{-1}$ \\
$\langle v_{\text{GSR}} \rangle$ & $-35.1_{-1.5}^{+1.5}$  & km\,s$^{-1}$ \\
$\sigma_{v_{\text{GSR}}}$ & $3.7_{-0.9}^{+1.5}$  & km\,s$^{-1}$\\
$\langle \text{[Fe/H]} \rangle$ & $-1.84_{-0.18}^{+0.16}$  & dex\\
$\sigma_{ \text{[Fe/H]}}$ & $0.37_{-0.13}^{+0.21}$  & dex\\
$\langle \text{[$\alpha$/Fe]} \rangle$ & $0.22_{-0.07}^{+0.07}$ &  dex\\
$\tau_\ast$ & $\gtrsim 12$ & Gyr\\ 
\tableline
\sidehead{Orbit}
$r_{\text{Gal}}$ & $14.2_{-0.4}^{+0.4}$ & kpc \\
$r_{\text{peri}}$ & $13.8_{-1.0}^{+0.5}$ & kpc \\
$r_{\text{apo}}$ & $18.4_{-2.9}^{+4.0}$ & kpc \\
$e$ & $0.14_{-0.04}^{+0.08}$ & \\
$E_{\text{tot}}$ & $-0.98_{-0.06}^{+0.06}$ & $10^5$\,km$^{2}$\,s$^{-2}$ \\
$L_{\text{Z}}$ & $1.57_{-0.12}^{+0.11}$ & $10^3$\,kpc\,km\,s$^{-1}$
\enddata
\end{deluxetable}

We perform a similar Gaussian fit to the radial velocities to resolve the line-of-sight velocity dispersion. 
Galactocentric radial velocities are used for this purpose, to remove any projection effects caused by the large spatial extent of the members. 
We add $0.5$~km\,s$^{-1}$ in quadrature to the individual radial velocity errors to account for systematic uncertainties in the wavelength calibration. 
Specter has a velocity dispersion of $3.7_{-0.9}^{+1.5}$~km\,s$^{-1}$ in its central region. 
This is similar to the dispersion measured in other dwarf streams like Orphan, Indus and Chenab \citep{Li2022b}. 
Specter has $f = (r_{\text{Gal}} - r_{\text{peri}})/(r_{\text{apo}} - r_{\text{peri}}) \approx 0.1$, matching the observed trend from \cite{Li2022b} that streams close to pericenter tend to have colder velocity dispersions, in contrast with theoretical predictions \citep{Helmi1999,Panithanpaisal2021}. 
This discrepancy could be a selection effect, particularly in the case of Specter, since it was detected due to the fast tangential velocity and cold radial velocity distribution of its members. 

\begin{figure}
    \centering 
    \includegraphics[width=\columnwidth]{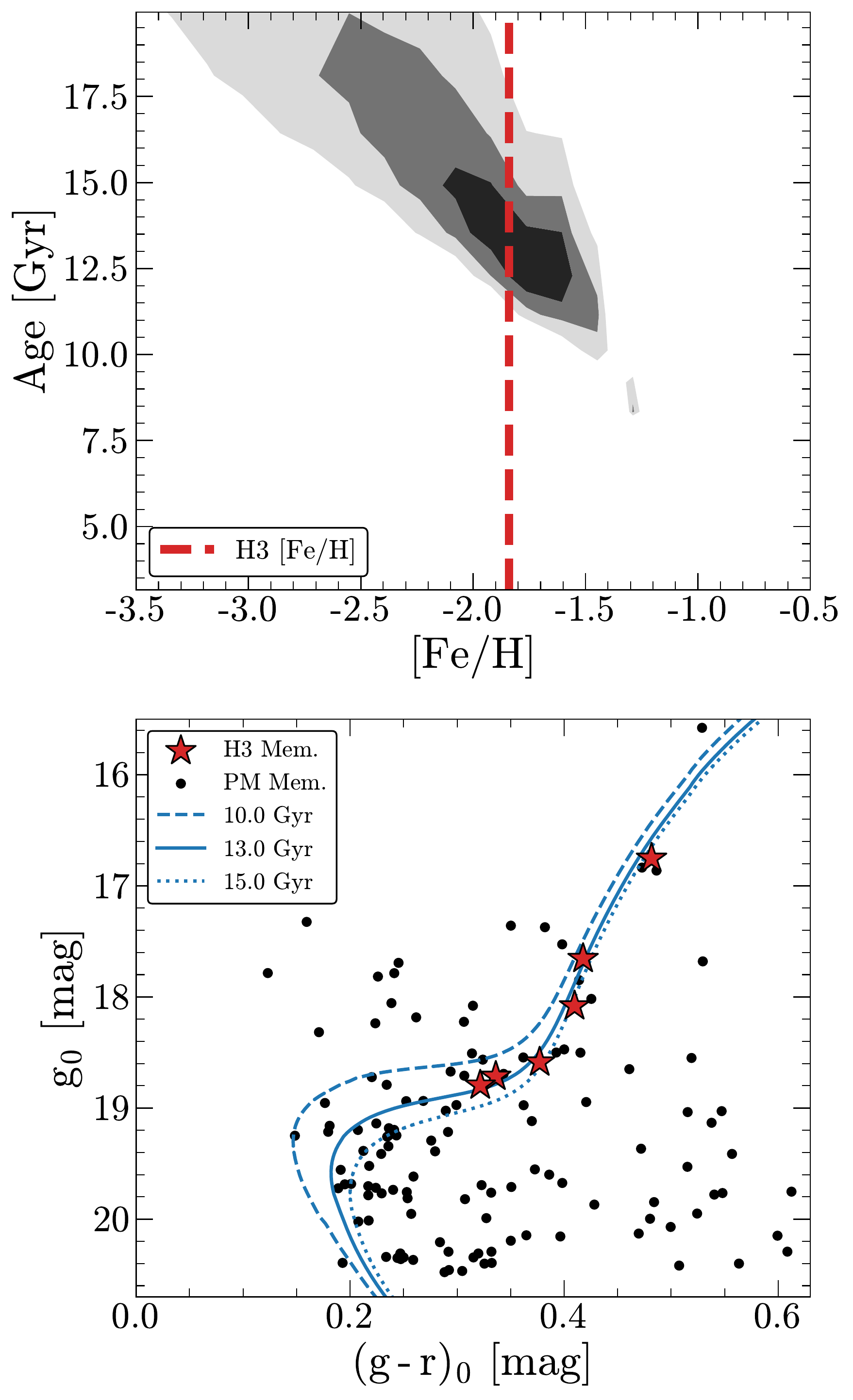}
    \caption{Top: $1-3\sigma$ confidence contours derived by fitting a grid of MIST isochrones to the six non-BHB spectroscopic members in Specter. The mean H3 spectroscopic metallicity is overlaid. Bottom: Pan-STARRS CMD of PM-selected members in the inner region of the stream, with the spectroscopic members highlighted as red stars. The overdensity of turnoff stars is best matched by a $\sim 13$~Gyr isochrone, and is inconsistent with an isochrone younger than $\sim 12$~Gyr at this metallicity.}
    \label{fig:cmdfit}
\end{figure}

\begin{figure*}
    \centering 
    \includegraphics[width=\textwidth]{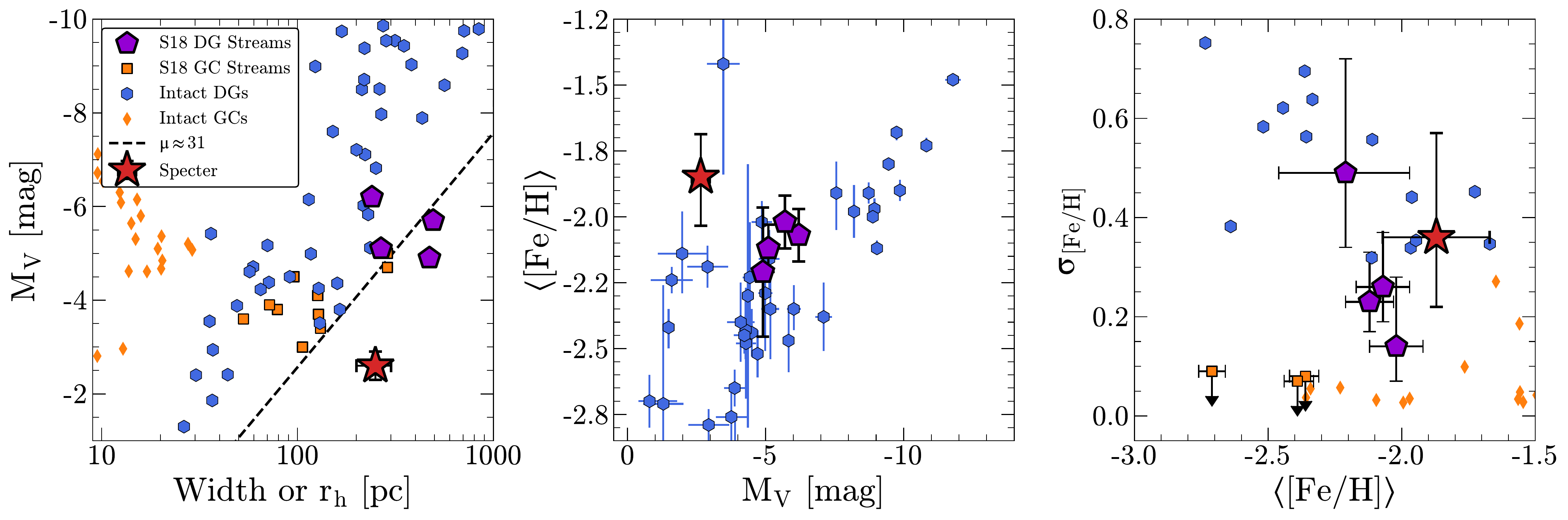}
    \caption{Comparing the luminosity, size, and metallicity of Specter to known streams and intact structures. Left: Stream widths (Gaussian $\sigma$) and luminosities for several DG and GC streams from \cite{Shipp2018}. We overlay intact (gravitationally bound) dwarfs and GCs, using the half-light radius as a proxy for characteristic size \citep{McConnachie2012,Baumgardt2020,Vasiliev2021}. Middle: Luminosity-metallicity relation for Specter, alongside an updated sample of intact dwarfs from \cite{McConnachie2012} and DG streams from \cite{Shipp2018}. Right: Mean metallicity and intrinsic metallicity spreads for streams from \cite{Ji2020}. We overlay intact dwarfs and GCs compiled by \cite{Willman2012}, omitting error bars for clarity.}
    \label{fig:stream_compare}
\end{figure*}

The H3 spectroscopic members have a median spectro-photometric age of $\sim 12$~Gyr as reported by \texttt{MINESweeper}, although these ages are affected by a Galactic model-dependent prior. 
To further characterize the stellar population in Specter, we fit the Pan-STARRS CMD of six spectroscopic members with a grid of MIST isochrones. 
The BHB member is excluded since models of BHB temperatures are quite uncertain, and will bias our result. 
We fix [$\alpha$/Fe]$=+0.3$ for the isochrones, and compute the total $\chi^2$ difference between the observed and isochrone-predicted $g-r$ color for these stars over a grid in age and metallicity. 
The resulting likelihood contours are illustrated in the top panel of Figure~\ref{fig:cmdfit}. 
The spectroscopic members strongly suggest an age older than $\gtrsim 12$~Gyr, particularly considering the mean spectroscopic metallicity of Specter. 
Furthermore, the larger PM-selected sample is also consistent with this isochrone, and disfavors younger ages  (Figure~\ref{fig:cmdfit}, bottom). 


Figure~\ref{fig:stream_compare} places Specter in the broader context of stellar streams and intact stellar populations around the Milky Way.
Specter's large metallicity spread is typical of dwarf galaxies around the Milky Way --- as shown in the right panel of Figure~\ref{fig:stream_compare} --- and argues against a globular cluster origin \citep{Kirby2011,Willman2012}. 
The most striking feature of Specter is its low luminosity, even lower than the globular cluster streams presented in \cite{Shipp2018}. 
Both the stream width and intrinsic metallicity spread of Specter suggest a dwarf galaxy origin for the stream. 
Our observations therefore indicate that Specter is a disrupted ultra-faint dwarf galaxy, comparable in stellar mass to \bootes{}~2, Carina~3, or Willman~1 \citep{Willman2005a,Walsh2007,Martin2008,Willman2011}. 
Specter has likely escaped detection thus far due to a combination of its low surface brightness (reducing its significance in photometric searches) and low star count (reducing its significance in 5D kinematic searches). 
Its detection required serendipitous H3 spectroscopy to identify the original comoving pair. 
The left panel of Figure~\ref{fig:stream_compare} shows that Specter is almost an order of magnitude wider than known intact dwarfs at similar total luminosity. 
One interpretation is that Specter's progenitor was structurally similar to Segue~1 or Willman~1, and the stream width has substantially broadened over time in the aspherical potential of the Milky Way. 
Specter is on a relatively polar orbit, a configuration that should promote rapid broadening of the stream \citep{Erkal2016}. 
However, past searches for intact dwarfs have been limited by a well-known surface brightness threshold of $\mu \sim 31$~mag\,arcsec$^{-2}$ \citep{Koposov2008,Tollerud2008,Walsh2009,Drlica-Wagner2021}, and the known population of intact dwarfs may not encompass all possible progenitors for Specter. 
For example, `stealth' galaxies inhabiting lower-mass dark matter halos could have escaped detection thus far, but could be plausible progenitors for Specter-like streams \citep{Bullock2010}. 

The middle panel of Figure~\ref{fig:stream_compare} shows that Specter lies above the luminosity-metallicity relation for intact dwarfs, joining Grus~1, Segue~2, and Willman~1 in this region of parameter space \citep{Willman2011,Kirby2013b,Walker2016}. 
This could imply a higher progenitor mass for Specter than is presently observed, perhaps due to tidal stripping or additional stream components hiding behind the Galactic disk. 
Alternatively, these discrepant dwarfs could reflect a metallicity floor for galaxy formation, at least in some special formation environments \citep{Simon2007,Rafelski2012,Kirby2013b}. 



\newpage

\subsection{Orbital Kinematics and Associations}

\begin{figure}
    \centering 
    \includegraphics[width=\columnwidth]{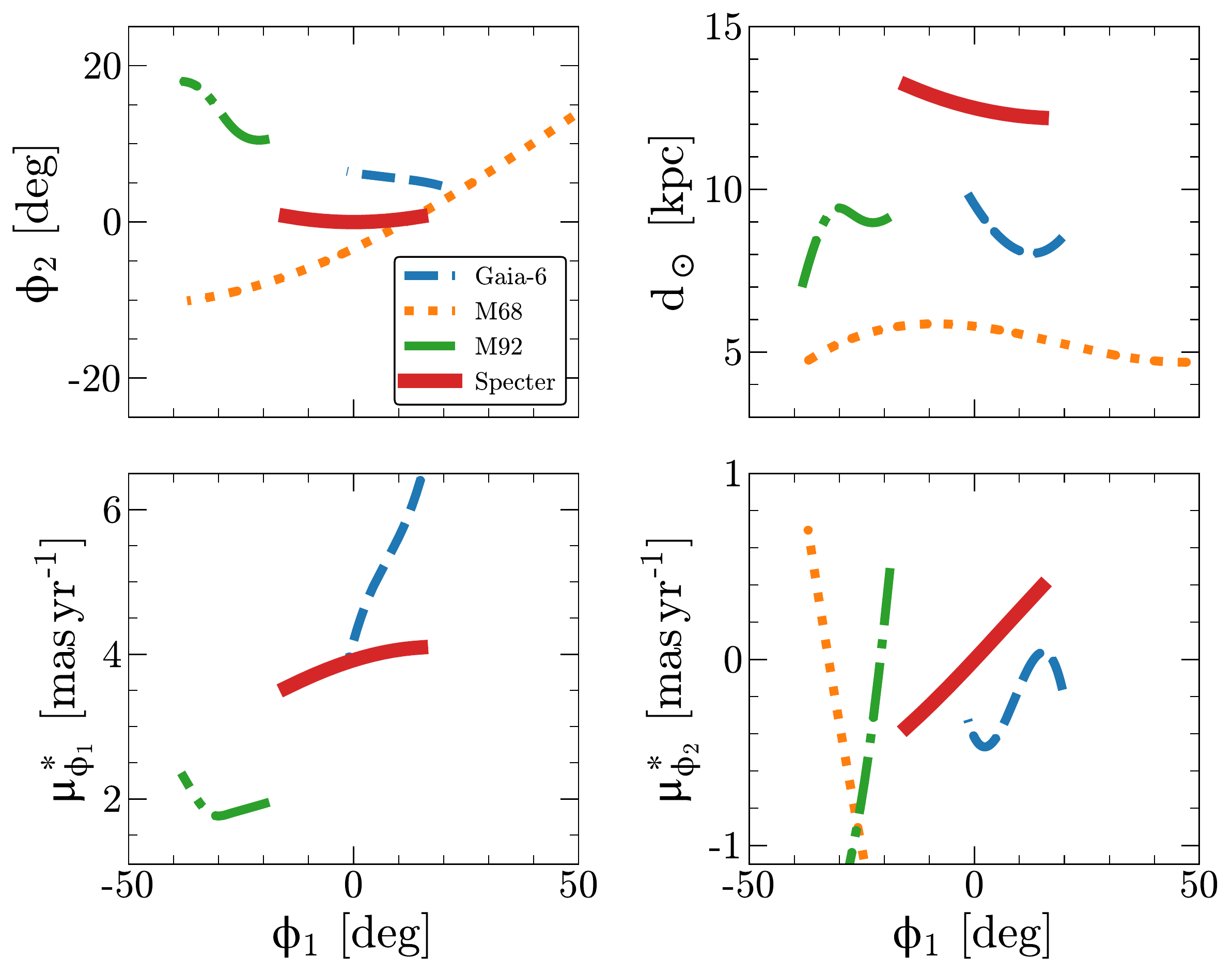}
    \caption{Comparing the observed orbital track, distance, and proper motions of Specter to some nearby known streams in the \texttt{galstreams} library.}
    \label{fig:galstreams_compare}
\end{figure}

Figure~\ref{fig:galstreams_compare} compares several observed properties of Specter to a subset of known streams compiled by \cite{Mateu2022}. 
In these panels, there is an apparent association between Specter and \textit{Gaia}-6 \citep{Ibata2021}. \textit{Gaia}-6 is $\approx$ 7 degrees offset from Specter with a similar size and orientation, and comparable proper motions. 
However, the proper motion gradient in \textit{Gaia}-6 is quite different than Specter, particularly in the stream-parallel component. 
\cite{Malhan2022b} report a median metallicity for \textit{Gaia}-6 of [Fe/H]$=-1.16$, which is more metal-rich than Specter, although the distribution of 10 member metallicities has a long tail towards the metal-poor end (K. Malhan, private communication). 
Additionally, \cite{Martin2022a} report a metallicity of [Fe/H]$=-1.50$ using narrow-band photometry of 12 members from the Pristine survey \citep{Starkenburg2017}. 
One possible interpretation is that \textit{Gaia}-6 either accreted with Specter, or represents a different orbital wrap of Specter \citep[e.g.,][]{Malhan2019}. 
Acquiring more velocities and metallicities in the \textit{Gaia}-6 stream will help resolve this question. 

\begin{figure*}
    \centering
    \includegraphics[width=\textwidth]{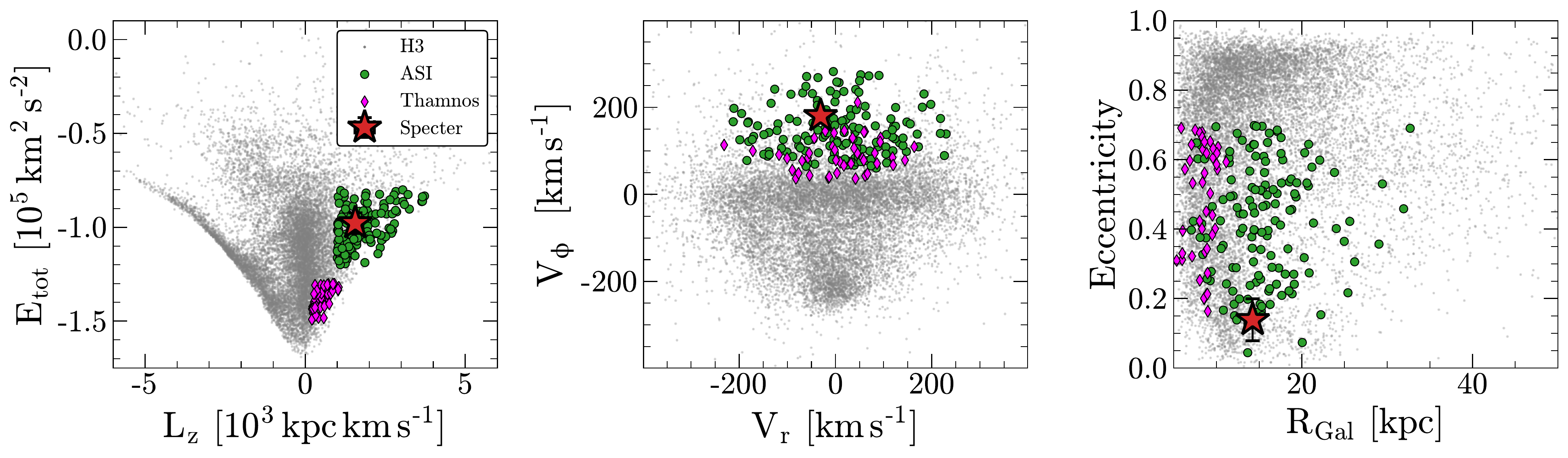}
    \caption{
    Specter in various integrals of motion. In grey we show the sample of giants from the H3 Survey. We highlight two known retrograde structures following the selections in \cite{Naidu2020}, Arjuna/Sequoia/I'itoi in green and Thamnos in magenta.}
    \label{fig:orbit_iom_summary}
\end{figure*}

We integrate Specter's orbit backward and forward in time using \texttt{gala} and compute integrals of motion for Specter following the methodology outlined in \cite{Naidu2020}, using the inverse-variance weighted mean kinematics of the H3 members. 
These are displayed in Figure~\ref{fig:orbit_iom_summary}, along with a sample of field giants from the H3 Survey for comparison. 
Specter is on a remarkably circular and retrograde orbit, with $e \approx 0.2$ and $L_{\text{Z}} \approx 1.6 \times 10^3\,\text{kpc\,km\,s}^{-1}$. 
We select and highlight two known retrograde structures following \cite{Naidu2020} --- Thamnos at low energies, and Arjuna/Sequoia/I'itoi at high energies \citep{Myeong2018c,Koppelman2019c,Myeong2019}. 
Specter is unlikely to be associated with Thamnos based on its total orbital energy alone. 
Additionally, the bulk of Thamnos debris lies closer to the Galactic centre around $R_{\text{Gal}} \approx 8$~kpc. 
Specter does overlap with Arjuna/Sequoia/I'itoi in energy and angular momentum space. 
However, Specter is more metal-poor than the bulk of Sequoia stars, more metal-rich than I'itoi, and generally more circular than the bulk of retrograde debris in H3 \citep{Naidu2020,Naidu2022}. 
We therefore note these associations, but do not find strong evidence that Specter is associated with any of these more massive disrupted dwarfs.
We also verify that no globular clusters lie along Specter's integrated orbit in 3D space, although this test is naturally sensitive to the assumed Galactic potential. 

\subsection{Detection Probability and Population}

Given our detection of Specter in the H3 Survey, we can roughly estimate how many low luminosity streams like Specter might exist at similar distances. 
In this case the `detection' signifies observing the original pair of comoving stars in the primary H3 sample. 
For a $\sim 2000\,M_\odot$ stellar population at 12.5~kpc, a \cite{Kroupa2001} IMF and our fiducial MIST isochrone predicts $\approx 13$ stars to fall within the H3 magnitude limits $15 \lesssim r \lesssim 18.5$. 
We perform a simple Monte Carlo test to estimate the probability of detecting a pair of stars from a theoretical Specter-like stellar population. 
We generate mock streams by randomly sampling pole coordinates uniformly on the sky, and create corresponding great circle coordinate frames. 
For each mock stream, we sample a Poisson distribution centered at $\langle N_{\text{true}} \rangle = 13$ and randomly place that number of stars on a $25^\circ \times 1^\circ$ region similar to Specter (distributed uniformly in $\phi_1$ and normally in $\phi_2$). 

Each mock stream is then `observed' by checking if any stars fall within the H3 tiles observed to date. If so, the stars are subjected to the sampling fraction (proportion of targets that get assigned fibers) for that tile. 
We repeat this mock stream generation and observation for 1000 trials, and use the fraction of trials in which at least 2 stars are `observed' as the probability of detecting a population with $N_{\text{true}}$ stars. 
By further repeating these trials for 1000 Poisson realizations of $N_{\text{true}}$, we get a distribution of representative detection probabilities. 
The median detection probability is $\approx 3\%$, which matches our na{\"i}ve expectation that the dominant contribution will be the $\approx 2.5\%$ survey footprint fraction of the sky. 
Put differently, since there are $\approx 13$ Specter stars within the H3 magnitude limit, there is a very high probability that a member pair will be observed if a field lands on the stream. 
Our Monte Carlo trial folds in the inherent stochasticity of the stellar population, as well as a more accurate model of the interaction between stream geometry, projected density, and the sparse survey footprint. 

Based on these trials, and the fact that we detected one such stream in H3, the expected number of true streams (at similar distance) can be estimated as the inverse of the detection probability. 
Taking the 16th and 84th quantiles of our Monte Carlo trials as a 1-sigma interval, we estimate 20--50 undetected Specter-like streams lying between $10-20$~kpc of the Sun. 
If we use this number to normalize a \cite{Navarro1997} number density profile and extrapolate to larger Galactocentric radii, we expect that hundreds of streams like Specter might reside within the Milky Way's virial radius \citep[e.g.,][]{Koposov2008,Drlica-Wagner2020}. 
Although our model is simplistic, we make a strong prediction that dozens more Specter-like streams await detection in future spectroscopic surveys like SDSS-V \citep{Kollmeier2017}, DESI \citep{DESICollaboration2016,AllendePrieto2020}, 4MOST \citep{deJong2019}, and WEAVE \citep{Dalton2020}.  

\section{Discussion}\label{sec:discuss}

We have presented the discovery and structural analysis of Specter, the least-luminous disrupted dwarf galaxy stream yet known in the Milky Way. 
It is remarkable that Specter was discovered using a single pair of metal-poor giants based on their proper motions and radial velocities, in an otherwise blind spectroscopic survey.
Specter joins a handful of objects that were detected as cold structures in spectroscopic surveys, a class originated by the Sagittarius dSph \citep{Ibata1994}. 
In fact, it is perhaps the first structure since Sagittarius to be revealed by radial velocity measurements themselves, as opposed to integrals of motion \citep[e.g.,][]{Tenachi2022}. 

Naturally, the distinction of Specter being the least-luminous dwarf galaxy stream assumes that we have detected the majority of its spatial extent. 
As shown in Figure~\ref{fig:selection}, below $\phi_1 \approx -25^\circ$ our kinematic selection increasingly fails to filter out the rising density of disk contaminants at lower Galactic latitudes. 
This leaves open the possibility that Specter has hidden components behind the disk, which could help to reconcile Specter's low luminosity for its metallicity (Figure~\ref{fig:stream_compare}, middle). 
However, the stellar density fits shown in Figure~\ref{fig:density_fit} exhibit a characteristic decrease in density on both sides of the stream, and our maps rule out a continued extent of the stream beyond $\phi_1 \gtrsim 15^\circ$. 
Furthermore, Specter would be less luminous than the dwarf galaxy streams in \cite{Shipp2018} even if its star count was inflated by a factor of five. 
For these reasons, we classify Specter as the least luminous dwarf galaxy stream known, and emphasize that the luminosity of its progenitor galaxy is relatively more uncertain.

A tantalizing possibility is that an intact dwarf analog to Specter's progenitor has not yet been observed due to the surface brightness limitations of current search techniques \citep[e.g.,][]{Tollerud2008}.
In particular, Specter's progenitor may have been an extended `stealth' dwarf galaxy that inhabited a low-mass dark matter halo \citep{Bullock2010}. 
The existence of such galaxies depends on the efficiency of galaxy formation at the lowest mass scales, and future deep surveys and kinematic searches may reveal intact analogs of a `stealth' progenitor for Specter. 
If instead Specter's progenitor was an ultrafaint dwarf like Segue~1 or Willman~1, then the stream's large $\approx 200$~pc width requires a dynamical explanation. 
Numerical simulations of the tidal disruption of various progenitor types may shed light on this problem --- probing whether such a short and wide stream (with no discernible progenitor today) can be produced by a Segue~1-like galaxy, or whether it suggests the gradual elongation of a more diffuse and stealthy progenitor. 

It is quite likely that Specter is the first known representative of a large class of disrupted dwarfs lurking below the detection threshold of current search techniques.
Our simple Monte Carlo trial of the H3 selection function implies that dozens more diffuse systems like Specter could reside at similar distances, and hundreds more might lie undiscovered throughout the Galaxy. 
This is further supported by recent discoveries of extremely diffuse debris using 6D kinematics \citep{Tenachi2022,Oria2022}.
Although most streams discovered by kinematic techniques do not have formal luminosity estimates, several are likely as ephemeral as Specter \citep[e.g.,][]{Ibata2019,Malhan2022b,Ji2022}. 
Our work motivates follow-up spectroscopy of these streams to ascertain if they might have dwarf galaxy progenitors. 

This work has demonstrated the power of spectroscopic surveys to catch extremely diffuse structures that would not form strong overdensities on the CMD or in 5D kinematics, but are revealed by their line-of-sight velocities. 
The detection of comoving pairs informs us about `where to look' in kinematic and chemical spaces, increasing the detection significance of otherwise sub-threshold structures. 
Future searches similar to those presented here, performed with upcoming large-scale spectroscopic surveys, are sure to unveil more hidden ghosts in the Galactic halo. 

\begin{acknowledgments}

We thank the anonymous referee for a thorough and constructive report that significantly improved our manuscript. VC gratefully acknowledges a Peirce Fellowship from Harvard University, and the hospitality of the Max Planck Institute for Astronomy, where a portion of this work was completed. 
We thank 
Gus Beane, 
Khyati Malhan, 
Hans-Walter Rix, 
Will Cerny, 
Nicolas Martin, 
and Zhen Yuan 
for insightful conversations. 
We are grateful to Alex Ji for sharing data from \cite{Ji2020}.  
CC and PC acknowledge support from NSF grant NSF AST-2107253. 
N.C. is supported by NSF grant AST-1812461.

Observations reported here were obtained at the MMT Observatory, a joint facility of the Smithsonian Institution and the University of Arizona. 
We are grateful to the Hectochelle operators at the MMT observatory for their tireless work and support throughout the H3 Survey. 
This work has made use of data from the European Space Agency (ESA) mission {\it Gaia} (\url{https://www.cosmos.esa.int/gaia}), processed by the {\it Gaia} Data Processing and Analysis Consortium (DPAC, \url{https://www.cosmos.esa.int/web/gaia/dpac/consortium}). Funding for the DPAC has been provided by national institutions, in particular the institutions participating in the {\it Gaia} Multilateral Agreement. 
This research has made extensive use of NASA's Astrophysics Data System Bibliographic Services.

\end{acknowledgments}

\facilities{
MMT, 
\textit{Gaia},
PS1
}

\software{
\texttt{numpy} \citep{Harris2020},
\texttt{scipy} \citep{Virtanen2020},
\texttt{astropy} \citep{Robitaille2013,Price-Whelan2018},
\texttt{matplotlib} \citep{Hunter2007},
\texttt{emcee} \citep{Foreman-Mackey2013,Foreman-Mackey2019},
\texttt{gala} \citep{gala,adrian_price_whelan_2020_4159870}
}

\appendix

\section{Significance of Detection in H3}\label{sec:detsig}

\begin{figure}
    \centering
    \includegraphics[width=\columnwidth]{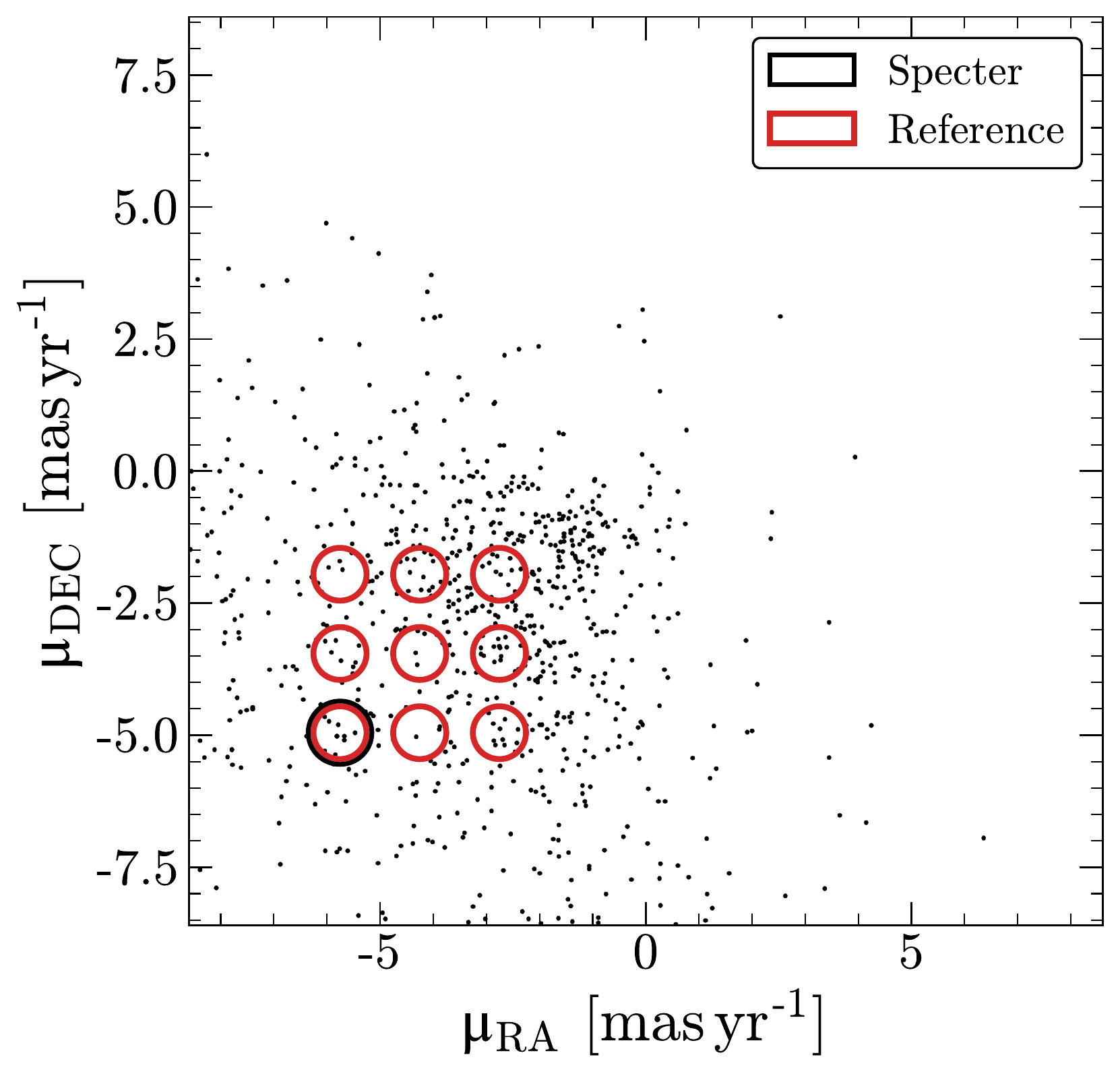}
    \includegraphics[width=\columnwidth]{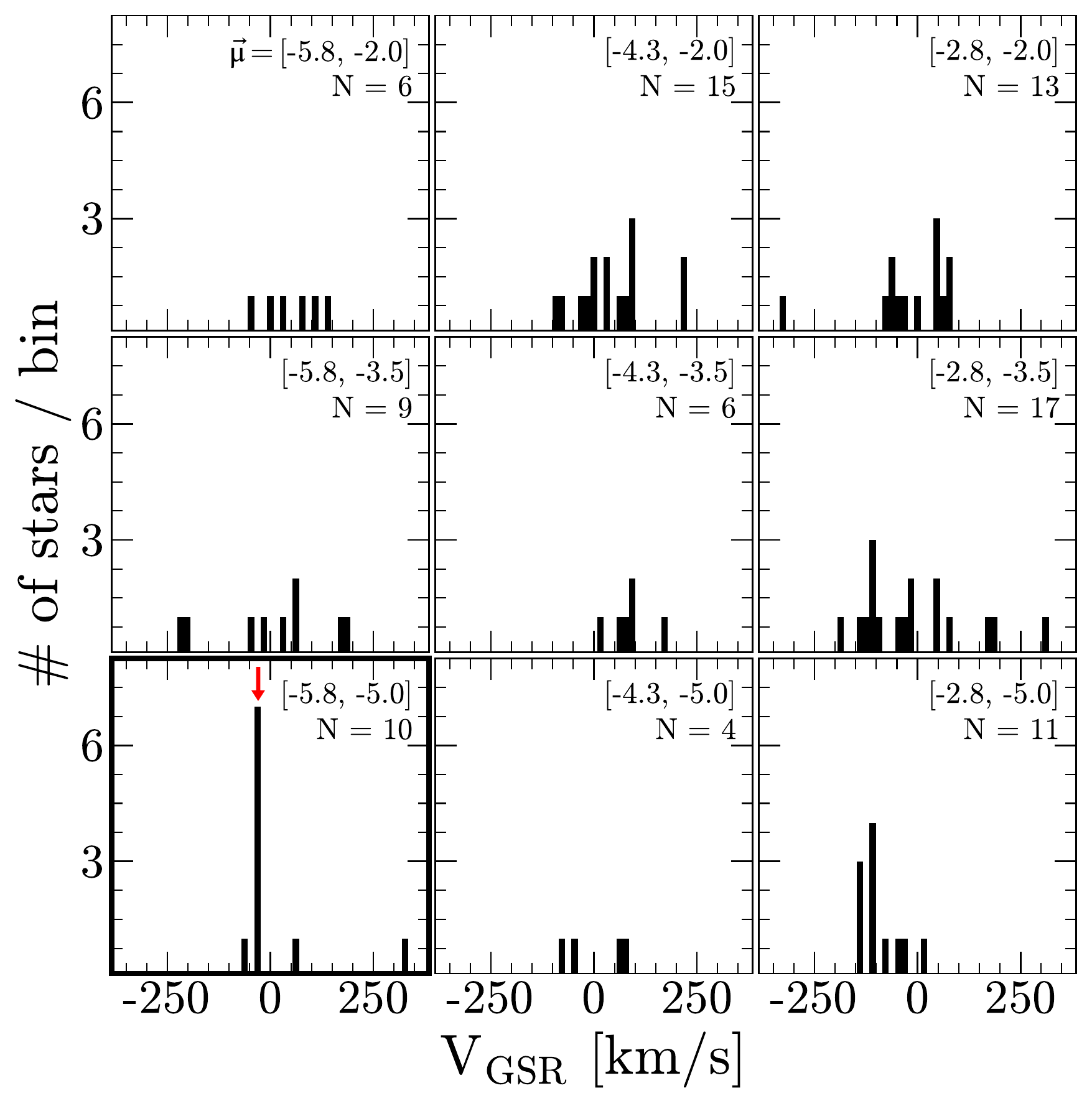}
    \caption{Demonstrating Specter's significant RV signal in the H3 dataset. Top: Equatorial PMs of CMD-selected giants in H3 within $25^\circ$ of Specter, with 9 selection regions overlaid near Specter's mean PM. Bottom: Galactocentric RV distributions in each PM selection region. Specter stands out as a sharp spike in the RV distribution, and there are no similar spurious spikes in other PM-selected regions.}
    \label{fig:pmtest}
\end{figure}

Our final sample of seven spectroscopic members from H3 has a non-trivial selection function, because two stars were observed in the regular survey, and five were subsequently followed up with PM-selected targeting. It is therefore prudent to ascertain whether targeting stars with a tight selection in PMs could spuriously produce a clustering in radial velocities. We perform a simple empirical test using H3 data for $\approx 20\,000$ giants with clean measurements. We select stars within $25^\circ$ of Specter and apply the CMD selection shown in Figure~\ref{fig:selection}, resulting in $\approx 1000$ giants whose proper motions are illustrated in the top panel of Figure~\ref{fig:pmtest}. We construct a 3x3 grid of proper motion selections with $0.5^\circ$ radii, with the bottom-left selection centered on the mean proper motion of Specter. For each selection, we plot the corresponding histogram of Galactocentric radial velocities in the bottom panel of Figure~\ref{fig:pmtest}. The histograms are computed over identical $[-350,350]$~km\,s$^{-1}$ ranges with a bin width of $15$~km\,s$^{-1}$.

This test demonstrates that Specter stands out as a prominent overdensity in radial velocities once the CMD and proper motion selection is applied. Furthermore, it illustrates that these selections are unlikely to spuriously produce such a sharp RV spike from field stars. Although our PM-selected follow up may over-represent the total number of stars with proper motions similar to Specter, it would not artificially enhance the concentrated distribution of RVs around Specter. Therefore, we conclude that Specter is significantly detected in the H3 spectroscopic dataset. The corresponding overdensities in the broader \textit{Gaia} dataset (Figure~\ref{fig:selection}) confirm that we have identified a bonafide coherent stellar population, rather than a chance projection of field stars. 

\section{Spectra of H3 Members}\label{sec:h3spec}

\begin{figure*}
    \centering
    \includegraphics[width=\textwidth]{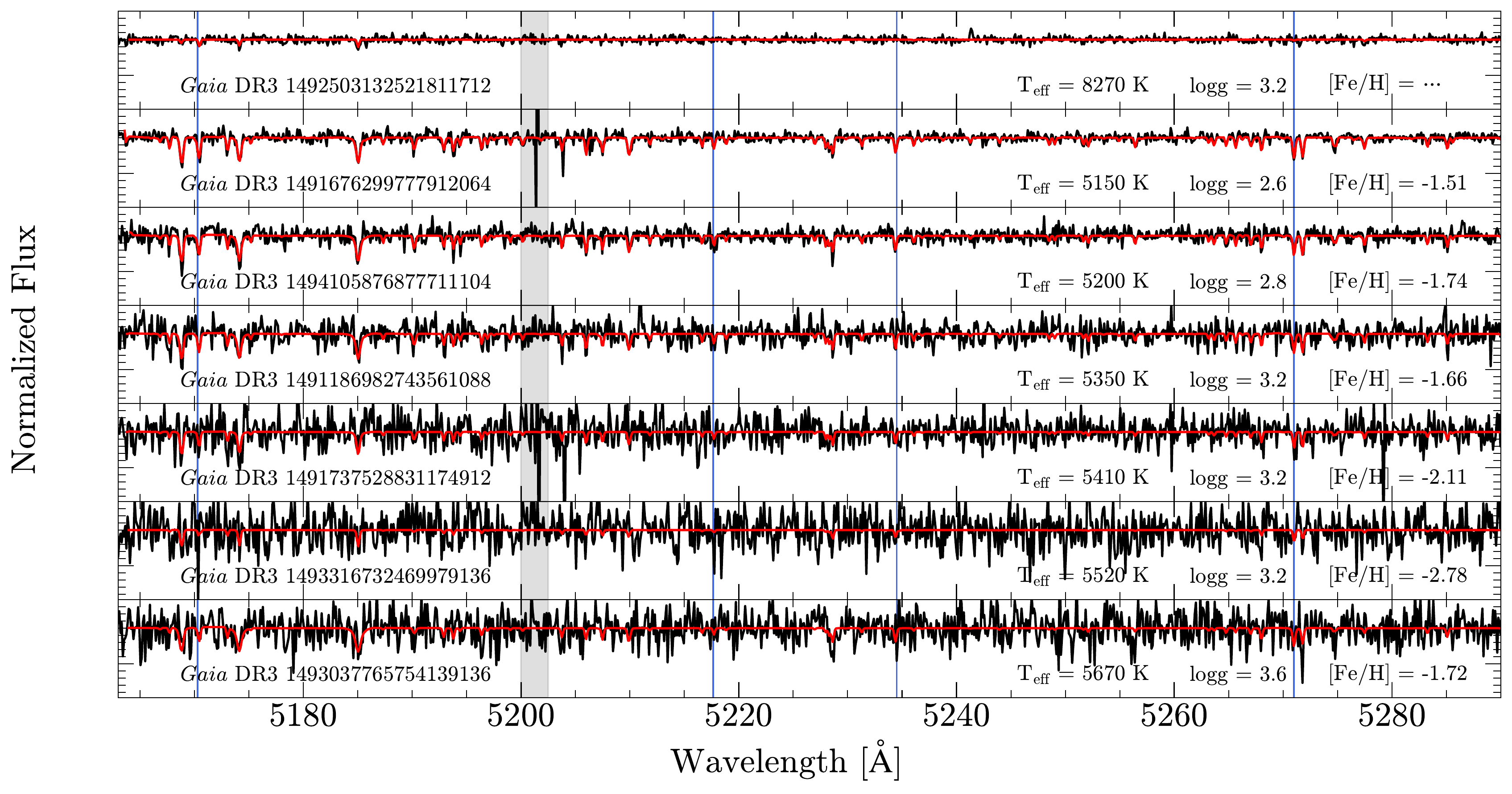}
    \includegraphics[width=0.8\textwidth]{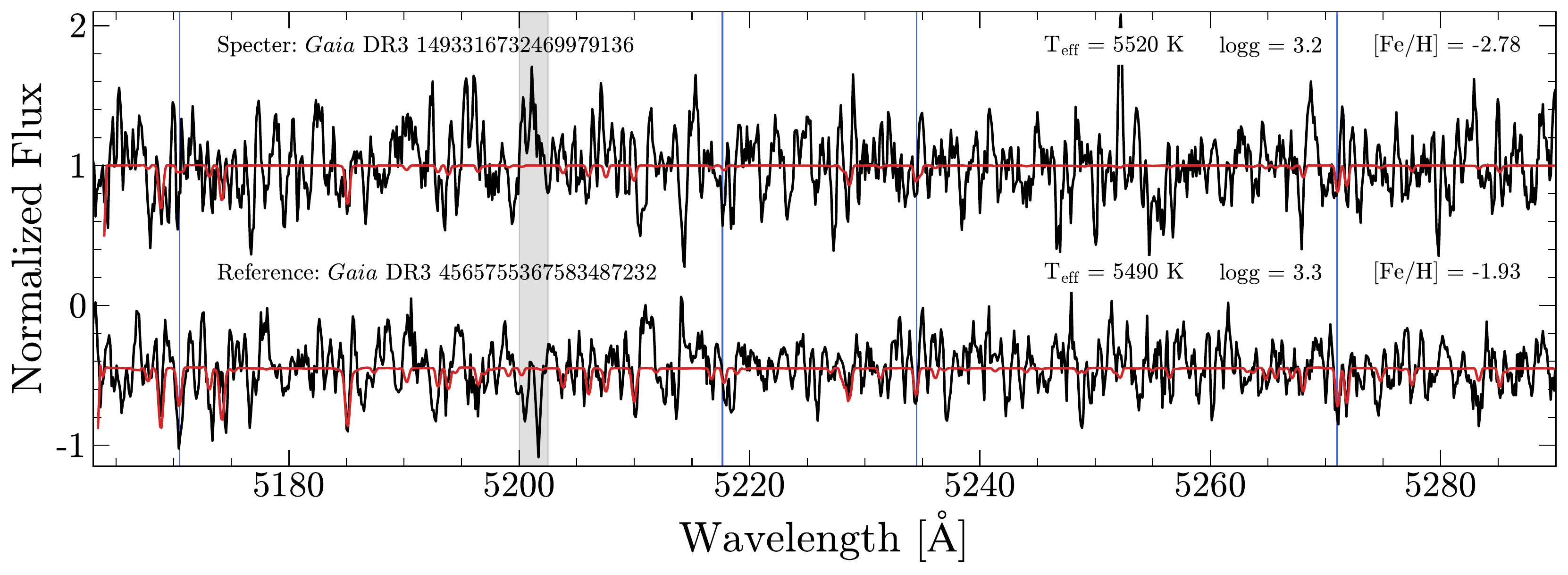}
    \caption{Top: H3 spectra for seven members in Specter, with the best-fitting \texttt{MINESweeper} model overlaid in red. The topmost star is the BHB, for which a reliable metallicity cannot be estimated. A masked telluric region is shown in grey, and we indicate strong Fe lines in blue. Bottom: Comparison between H3 spectra for the most metal-poor member in Specter, and a reference star at higher metallicity selected from the H3 catalog to have similar temperature, surface gravity, and spectral signal-to-noise ratio. Both spectra are smoothed with a 5 pixel boxcar. Note in particular the Fe line at $\approx 5170$\,\AA~that is much weaker in the Specter member than the reference star. We emphasize that the full-spectrum-fitting \texttt{MINESweeper} routine picks up on aggregate spectral details that are difficult to discern by eye. Regardless, this comparison reassures us about this member's metallicity.}
    \label{fig:h3spectra}
\end{figure*}

We illustrate the $R \approx 32\,000$ H3 spectra for our seven spectroscopic members in Figure~\ref{fig:h3spectra}, arranged in the same brightness ordering as Table~\ref{tab:spec_mem}. We display the best-fitting \texttt{MINESweeper} models and stellar parameters.

The most metal-poor star in our sample, \textit{Gaia} DR3 1493316732469979136 deserves further comment since its metallicity is a key part of our argument for Specter's dwarf galaxy origin ($\S$\ref{sec:stellpop}). This star has a relatively low SNR in our sample. However, by comparing it to the next faintest star in Specter (at similar SNR and stellar parameters, bottom panel of Figure~\ref{fig:h3spectra}), the lack of strong iron lines already suggests [Fe/H]~$\lesssim -2$ for this star. We have verified that \texttt{MINESweeper} derives the correct RV for this star by re-running the fit with a broader RV prior. The RV is well-measured, with a sharp Gaussian posterior distribution. We also ran a \texttt{MINESweeper} fit exluding the H3 spectrum, utilizing only the parallax and broadband optical-IR photometry. This photometry-only fit returns [Fe/H]~$= -3.0 \pm 0.5$, excluding the mean metallicity of Specter at $\sim 2.5\sigma$. Based on these lines of evidence, we argue that \textit{Gaia} DR3 1493316732469979136  is indeed metal-poor.

\clearpage

\bibliography{library}
\bibliographystyle{aasjournal}

\end{document}